\documentclass[useAMS,usenatbib]{mn2e}

\usepackage{psfig}
\usepackage{amssymb}    % additional math symbols such as 'similar less than'
\usepackage{ulem}

\newcommand{\kms}{km~s$^{-1}$}

\newcommand{\hb}{H$\beta$}				% H beta
\newcommand{\mgb}{Mg$\,b$}				% Mg b
\newcommand{\oiii}{[O{\small III}]}			% [OIII]

\newcommand{\sz}{$\sigma_z$}
\newcommand{\sr}{$\sigma_R$}
\newcommand{\st}{$\sigma_\phi$}
\newcommand{\szsr}{$\sigma_z/\sigma_R$}
\newcommand{\ages}{age-$\sigma$}

\title[Disk heating across the Hubble sequence]{Disk heating
    agents across the Hubble sequence\thanks{Based on observations made
      with ESO Telescopes at the La Silla Observatory under programmes
      074.B-0550(A) \& 078.B-0152(A).}}

\author[J.Gerssen \& K. Shapiro Griffin]
{J.~Gerssen$^{1}$\thanks{E-mail: jgerssen@aip.de}, K.~Shapiro Griffin$^{2}$ \\
$^{1}$Leibniz-Institut f\"ur Astrophysik Potsdam, An der Sternwarte 16, 14482 Potsdam, Germany.\\
$^{2}$Space Sciences Research Group, Northrop Grumman Aerospace Systems, Redondo Beach, CA, USA}

\begin{document}

%\date{}
%\pagerange{\pageref{firstpage}--\pageref{lastpage}} \pubyear{2002}

\maketitle

\label{firstpage}

\begin{abstract}
We measure the shape of the velocity ellipsoid in two late-type spiral
galaxies (Hubble types Sc and Scd) and combine these results with our
previous analyses of six early-type spirals (Sa to Sbc) to probe the
relation between galaxy morphology and the ratio of the vertical and
radial dispersions.  We confirm at much higher significance (99.9
percent) our prior detection of a tight correlation between these
quantities.  We explore the trends of the magnitude and shape of the
velocity ellipsoid axes with galaxy properties (colour, gas surface mass
density, and spiral arm structure).  The observed relationships allow
for an observational identification of the radial and vertical disk
heating agents in external disk galaxies.
\end{abstract}

\begin{keywords}
galaxies: kinematics and dynamics -- galaxies: individual: NGC~2280, NGC~3810
-- galaxies: structure -- galaxies: fundamental parameters
\end{keywords}

\section{Introduction}
\label{sec:intro}

More than half of the stellar mass in the local Universe is observed to
reside in disk galaxies (\citealp{Driver2007, Weinzirl2009}), yet the
evolutionary history of these systems remains poorly understood.
Simulations of galaxy formation tend to produce spiral galaxies with
most of their stellar mass fraction in a bulge component, and a direct
comparison between observations and these simulations is usually not
straightforward \citep{Scannapieco2010}.  However, increasingly powerful
simulations of disk galaxy formation are reaching a stage where internal
dynamical properties, such as rotational and random velocities, can also
be investigated and compared to observations \citep{Scannapieco2011}.
These velocity distribution functions are fossil relics of the history
of gravitational forces acting on stars and therefore of the
evolutionary histories of spiral galaxies; for this reason, measurement
of the stellar velocity distributions in local spiral galaxies is an
important tool for probing the evolution of these systems.

From observations in the solar neighbourhood it has been shown that the
random motions of thin disk stars correlates with their ages, the
\ages\ relation (\citealp{Wielen1977, Binney2000}).  This relation is
most commonly parametrized as a power law of the form $\sigma \propto
t^\alpha$, with existing measurements of $\alpha$ spanning the range of
$0.2-0.5$ (see e.g. \citealp{Nordstrom2004} for results from the
Geneva-Copenhagen survey).  However, the \ages\ relation has also been
interpreted as saturating at a constant dispersion after $\sim5$~Gyr
(e.g. \citealt{Carlberg1985, Seabroke2007, Soubiran2008}) or as several
discrete increases in velocity dispersion (\citealp{Quillen2001}).  As a
result of the observational uncertainty, it is not straightforward to
identify the disk heating mechanism responsible for the rise in velocity
dispersions.

Additional constraints on the disk heating mechanism are provided by the
three-dimensional distribution of stellar velocity dispersions, whose
magnitude and direction designate the axes of an ellipsoid; this
trivariate Gaussian function was originally proposed by
\citet{Schwarzschild1907}.  The so-called velocity ellipsoid is
parameterized by its two axis ratios, $\sigma_\phi / \sigma_R$ and
$\sigma_z / \sigma_R$.  In an axisymmetric disk with stellar orbits not
too far from circular (the epicycle approximation), $\sigma_\phi /
\sigma_R$ depends only on the circular velocity and not on any disk
heating mechanism. Measurements of the ratio $\sigma_z / \sigma_R$ can
thus be used to probe the velocity distribution and therefore to
constrain the heating processes in galactic disks.

In the Milky Way, the \ages\ relation and velocity ellipsoid shape
(\szsr\ = 0.5) have been studied extensively, and a number of heating
mechanisms have been proposed to simultaneously explain the
observations.  These include encounters with giant molecular clouds
(GMCs, \citealt{Spitzer1951, Lacey1984, Lacey1985}), perturbations from
irregular and transient spiral structure \citealt{Barbanis1967,
  Jenkins1990, Fuchs2001, Minchev2006}), perturbations from stellar bars
\citep{Saha2010}, dissolution of young stellar clusters
\citep{Kroupa2002}, scattering by dark halo objects or globular clusters
\citep{Hanninen2002, vandePutte2009}, and disturbances by satellite
galaxies or minor mergers \citep{Velazquez1999}.
This plethora of disk heating theories reflects the
limited constraints provided by measurements in a single galaxy.
However, many of these theories are implicitly dependent on galaxy
morphology; progress towards an understanding of secular evolution in
disk galaxies therefore critically depends on data from external
galaxies spanning a range of Hubble types.

The numerical study of \citet{Jenkins1990} is one of the few that can be
directly compared with observations of disk heating in external
galaxies.  This theory combines two disk heating mechanisms, spiral
transients and GMCs, in what appears to be the simplest explanation that
is consistent with most of the observations.
By varying the relative importance of the two mechanisms they showed
that disk heating, as quantified by the ratio $\sigma_z / \sigma_R$,
decreases with the increase of spiral strength, e.g. along the Hubble
sequence (but see also \citealt{sellwood2008}).

We have developed a technique \citep{Gerssen1997} to measure the
velocity dispersion ratio in galaxies from spectra obtained along the
major and minor axis.  We applied this technique to six disk galaxies
spanning a range in Hubble type from Sa to Sbc (\citealt{Gerssen1997,
  Gerssen2000, Shapiro2003}).  In each system we were able to constrain
$\sigma_z / \sigma_R$ by averaging measurements over radii in the range
of one to two disk scale lengths.  While the uncertainties on these
ratios are large for individual galaxies, there appears to be a trend
with Hubble type that is consistent with the predictions of
\citet{Jenkins1990}.  However, the applicability of this trend to
later-type systems remained unclear \citep{Shapiro2003}.

In this paper, we address this issue by directly measuring the velocity
ellipsoid ratios in two late-type disk galaxies: NGC~3810 (Sc) and
NGC~2280 (Scd).  Section~\ref{s:Obs} presents our observations of these
galaxies.  Section~\ref{s:Analysis} describes our kinematic extraction
and modeling procedure.  Section~\ref{sec:results} presents our derived
velocity ellipsoid ratios and combines these results with our previous
work on early-type disks to examine the ratio of vertical to radial
velocity dispersions across the Hubble sequence.  The new results are
consistent with the trend observed earlier and suggest that the relative
effectiveness of vertical to radial secular evolution processes
diminishes as a function of galaxy morphology.  We additionally combine
our data with literature measurements of the velocity ellipsoid in
elliptical and lenticular galaxies \citep{sauronX} to reveal intriguing
trends across all galaxy morphologies.  In Section~\ref{sec:discussion},
we use our results to identify heating agents at work in spiral
galaxies, and in Section~\ref{sec:conclu}, we conclude.

\section{Observations \& data reduction}
\label{s:Obs}

\begin{table}
\begin{center}
    \caption{Galaxy properties}
    \begin{tabular}{@{}lccccc@{}}
    \hline
    Name  & Type & Redshift  & $i$  & Scale length &  Reference \\
          &      &  (km s$^{-1}$) &   (deg) &  (arcsec) &        \\
    \hline
    NGC~2280  & Scd & 1899   &     64      &   19     & 1, 2    \\
    NGC~3810  & Sc  &  993   &     32      &   16     & 1, 3    \\
    \hline
    \end{tabular}
    \label{t:properties}
\end{center}
\vspace{-0.25cm}
1) Redshifts and morphological classifications are from NED. \\
2) Scale length and inclination are derived in section~\ref{s:phot}. \\
3) Scale length and inclination are from \citet{knapen2003}. \\
\end{table}

\begin{table}
  \centering
    \caption{Observing log}
    \begin{tabular}{@{}lccc@{}}
    \hline
    Galaxy & Position angle & T$_{exp}$  & Date\\
           &   (deg)        & (min) & \\
    \hline
    NGC~2280 & & & \\
    \ \ \ Major axis &  -22 & 156 & Feb 2005 \\
    \ \ \ Minor axis & -112 & 160 & Nov 2006 \\
    NGC~3810 & & & \\
    \ \ \ Major axis &   25 & 210 & Feb 2005 \\
    \ \ \ Minor axis &  115 & 180 & Feb 2005 \\
    \hline
    \end{tabular}
    \label{t:obslog}
\end{table}

\subsection{Spectroscopy}

To probe the shape of the velocity ellipsoid at the end of the Hubble
sequence, we obtained long-slit spectra along the major and minor axes
of two late-type spiral galaxies.  The galaxy properties are listed in
Table~\ref{t:properties}, and the observing log is given in
Table~\ref{t:obslog}.  All spectra were obtained with the EMMI
spectrograph on the NTT during two runs in February 2005 and November
2006.  We used the red medium dispersion grating (\#6) to study the
wavelength range 4800 - 5500 \AA, which covers key interstellar emission
(\oiii, \hb) and stellar absorption (\mgb, several Fe) features.  The
slit length was 300\arcsec, and the slit width was 1\arcsec, resulting
in a spectral resolution of 0.4 \AA\ ($\sim 23$ \kms). Measurements of
the widths of unresolved arclines in our calibration spectra are
consistent with this value.

Observations were taken in exposures of either 1200s or 2400s, which
were interspersed with calibration exposures of a He+Ar lamp.  In
addition, we obtained spectra of several stars of spectral types G0 III
through K3 III, for use in extracting the stellar kinematics in our
sample galaxies (\S~\ref{StellarKin}); these stars are the dominant
component in the \mgb\ and Fe stellar absorption features of spiral
galaxies.  To distribute the light from the template stars relatively
uniformly on the detector and to prevent overexposure, the telescope was
defocused during the observations of these stars.

Preliminary reduction steps (bias subtraction, flat-fielding) were done
with our own IDL scripts, customized for EMMI data, and subsequent
reduction steps (wavelength calibration, coadding individual exposures,
extraction of 1-dimensional stellar spectra) were performed using
standard IRAF\footnote{IRAF is distributed by the National Optical
  Astronomy Observatory, which is operated by the Association of
  Universities for Research in Astronomy (AURA) under cooperative
  agreement with the National Science Foundation.}  packages.
For the kinematic analysis, we interpolate the spectral axis to
logarithmic units in wavelength (linear in velocity) and spatially bin
each galaxy spectrum to a minimum signal-to-noise (S/N) ratio of 10.

\subsection{Photometry}
\label{s:phot}

Since the quantities of interest are properties of the galaxy disks and
not of their bulges, we use broad-band images to measure the radii at
which the disk dominates the stellar surface brightness.  At these
radii, the derived kinematics probe the disk stars and not the bulge
and/or bar; we therefore limit our analysis (see \S~\ref{Modeling}) to
these radii.

\begin{figure}
   \centerline{\psfig{figure=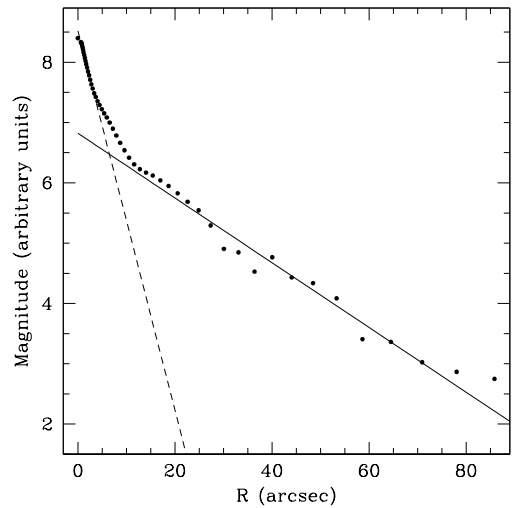,width=0.4\textwidth}}
   \caption{ The radial surface brightness profile of NGC~2280 derived
     from an archival {\it H}-band image obtained with the CTIO 1.5m
     telescope \citep{Eskridge2002}.  The best-fit exponential model to
     the outer disk (solid line) has an exponential scale length of
     19\arcsec. The dashed line is a guide (not a fit) delineating the
     bulge component. The presence of a third component, perhaps a small
     bar, can be inferred from the excess of light between 5 and 10
     arcsec.
\label{f:fig1}}
\end{figure}

For NGC~3810, \citet{knapen2003} have measured the disk scale length to
be $16$\arcsec\ and the disk to dominate the galaxy light at
$r>5$\arcsec.  For NGC~2280, we use an archival $H$-band image
\citep{Eskridge2002}, obtained with the CTIO 1.5m telescope and
available via the NED extragalactic database.  
We fit this image with the IRAF task ELLIPSE and derive an inclination
of 64 degrees (from the best-fit axis ratios) and the surface brightness
profile shown in Figure~\ref{f:fig1}.  The best-fit exponential disk
model to this profile has a scale length of $19$\arcsec\ and dominates
the galaxy light at $r>10$\arcsec.  
(Using a $K$-band image, \citet{Kassin2006} derive a scale length of 27
arcsec.)
The radial surface brightness profile (Figure~1) hints at the presence
of a small bar in NGC~2280.  Between the linear behaviour of the surface
brightness profile at large radii (the exponential disk) and the inner
bulge profile, there is a distinct third component. This component
dominates the surface brightness profiles at radii of 5-10\arcsec; the
combined shape of the surface brightness profile and flat minor axis
velocity dispersions (see \S~\ref{sec:n2280}) at these radii are
consistent with the presence of a bar (e.g. ~\citealt{Prieto2001,
  Gerssen2003}).

\section{Analysis}
\label{s:Analysis}

\subsection{Stellar and Gas Kinematics}
\label{StellarKin}

For each binned spectrum, we extract the stellar kinematics using the
{\it ppxf} method developed by \citet{cap04}.  We modeled the absorption
line kinematics from our galaxy spectra as convolutions of a template
stellar spectrum with line-of-sight velocity distributions (LOSVD); the
observation of the template stars with the same instrumental set-up as
the galaxies allows the instrumental broadening to drop out during this
fitting.  Given the relatively low S/N of our data, we did not attempt
to fit higher order moments of the LOSVD (e.g. $h_3, h_4$, etc) and
instead assumed the LOSVD to be Gaussian.  Since the outer disks of
late-type galaxies are faint and the spectra are dominated by sky
emission, we employ the method of \citet{Weijmans2009}, in which a
median sky spectrum taken from the edge of our long-slit is included as
a template in the fitting procedure.  We extract the stellar kinematics
using several GIII and KIII templates.  An optimal combination of these
templates did not significantly improve the results, so we chose the
K1III giant HD114971, which yields marginally better $\chi^2$, to
extract the stellar kinematics in both galaxies.

\begin{figure*}
   \centerline{\psfig{figure=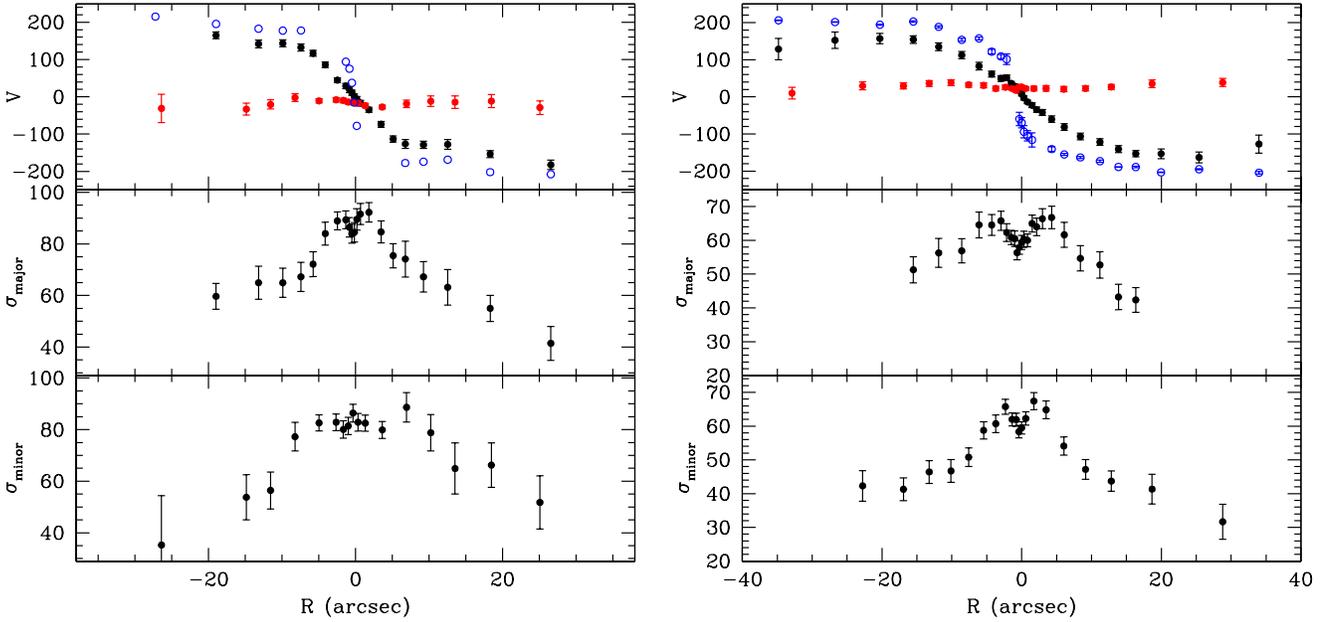,width=1.0\textwidth}}
   \caption{Observed kinematics of the two late-type spiral galaxies,
     {\bf left:} NGC~2280 (type Scd) and {\bf right:} NGC~3810 (type
     Sc).  In all panels the kinematic information is given in \kms.
     Filled circles represent major and minor axis stellar data; open
     circles represent gas data.  Error bars for the stellar velocities
     are shown but are generally smaller than the plot symbols.  The top
     panel for each galaxy shows the gas major axis rotation curve (blue
     open circles), the stellar major axis rotation curve (black filled
     circles), and the stellar minor axis velocities overplotted (red
     filled circles, roughly constant at 0 \kms).  The galaxy systemic
     velocities have been subtracted.  The lower two panels for each
     galaxy show the major and minor axis stellar velocity dispersions
     respectively.  The abscissa is the radius in the disk, which, for
     the minor axis, was obtained through deprojection.
   \label{f:fig2}}
\end{figure*}

The derived stellar velocities and velocity dispersions along the
major and minor axes of our sample galaxies are presented in
Figure~\ref{f:fig2}.  In these plots, the abscissa is the radius in
the plane of the disk, which for the minor axes was obtained through
deprojection.  The velocity dispersion profiles for both galaxies show
central drops.  This is a fairly common phenomenon
(e.g. \citealt{Comeron2008}) and is usually attributed to a central disk or
nuclear bar.  These central velocity dispersion drops do not affect
the results presented in this paper as their influence is confined to
radii that we exclude from our analysis.

To probe the circular velocity in our target galaxies we use the
\oiii\ emission line at 5007~\AA.  We fit this line in each spatial
bin with a Gaussian profile to derive the best-fit centroids.  As the
lines are unresolved in our data (width $<23$ \kms) we do not attempt
to quantify the gas velocity dispersions.  In the following, we assume
the measured gas velocities to be a good approximation of the circular
velocity in our galaxies.

\subsection{Modeling}
\label{Modeling}

To measure the velocity ellipsoid in our late-type galaxies, we model
the stellar and gas kinematics using the method developed by
\citet{Gerssen1997} and subsequently applied to a larger range of
morphological types by \citet{Gerssen2000} and \citet{Shapiro2003}.  

Briefly, the line-of-sight (los) kinematics along the major and minor
axes of intermediate inclination galaxies probe different projections of
the velocity ellipsoid, namely \st\ and \sr\ along the major axis and
\sr\ and \sz\ along the minor axis:

\begin{equation}
\sigma^2_{\rm major} = \sigma_R^2 \sin^2 i + \sigma_z^2 \cos^2 i
\end{equation}

\begin{equation}
\sigma^2_{\rm minor} = \sigma_\theta^2 \sin^2 i + \sigma_z^2 \cos^2 i
\end{equation}

The epicycle approximation (e.g. Eq. 2 in \citealt{Shapiro2003}) 
for stellar orbits in a disk yields \st/\sr\ as a function of only the
circular velocity $V_c(R)$ of the system.  Assuming axisymmetry,
measurements of $\sigma_{\rm los}$ along the major and minor axes and of
$V_c$ along the major axis fully constrain the shape of the velocity
ellipsoid.  Including the stellar rotation curve $\overline{V}(R)$ and
modeling the asymmetric drift as a function of only the disk scale
length and \sr\ (i.e. assuming no tilt $\sigma_{Rz}$ to the velocity
ellipsoid; see \citealt{Gerssen1997}) then overdetermines the system and
provides an additional constraint to the model:

\begin{equation}
\overline{V}^2 = V_c^2 - 
\sigma_R^2 \left[\frac{R}{h} - R \frac{\partial}{\partial R}
\ln(\sigma_R^2) - \frac{1}{2} + \frac{R}{2V_c} \frac{\partial
V_c}{\partial R} \right] 
\end{equation}

For a more detailed description of this method the reader is
referred to \citet{Shapiro2003}.
In practice, the low intrinsic luminosity and small velocity dispersion
of our late-type galaxies (compared to the early-type disks studied by
\citealt{Gerssen1997, Gerssen2000, Shapiro2003}) resulted in only a few
data points sampling the radial behavior of the kinematic profiles.
This paucity of data points forced us to include the emission line
kinematics directly in the modelling rather than use them as an
independent check on the best-fit results in our previous analyses.

The four observables ($\sigma_{\rm major}, \sigma_{\rm minor}, V_c,$
and $\overline{V}$) are simultaneously fit to produce \sz$(R)$,
\sr$(R)$, and $V_c(R)$.  The velocity dispersions as a
function of radius are assumed to be exponentially declining with
identical scalelength $h_{kin}$, parametrized by $\sigma_{z,0}$,
$\sigma_{R,0}$, and $h_{kin}$, and the circular velocity is assumed to
be a power law with radius, described by $V_{c,0}$ and power-law index
$\alpha$, for a total of five fitted parameters.

\begin{figure*}
   \centerline{\psfig{figure=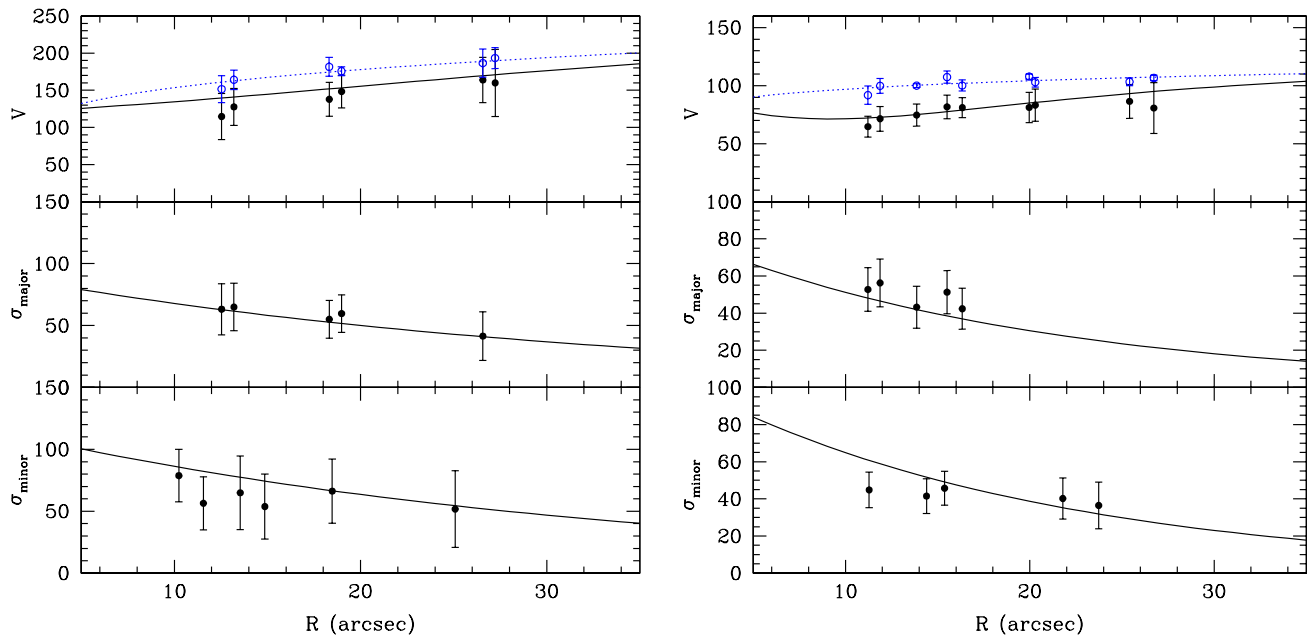,width=1.0\textwidth}}
   \caption{Similar to Fig.~2 but showing the observed kinematic
     quantities at absolute radii and excluding data contaminated by
     bulge light, {\bf left:} NGC~2280 and {\bf right:} NGC~3810.  All
     kinematic quantities (filled circles: stars, open circles: gas) are
     in \kms.  The best-fit models to the observed stellar kinematics
     (Eq. 3, 1 and 2 respectively from top to bottom) are shown as solid
     lines.  Unlike our previous study \citep{Shapiro2003}, where the
     gas velocities were used as a consistency check on the circular
     velocities (dotted lines), we include the gas velocities directly
     in our fitting procedure to better constrain the velocity
     ellipsoids in these later-type disk galaxies.
   \label{f:bestfit}}
\end{figure*}

As our data set is rather small with fairly large error bars on each
point, we employ two parameter space minimization techniques: the
Levenberg-Marquardt method and a simulated annealing method.  In both
cases we follow the implementation of these method as described in
\citet{Press1992}. Both methods fit our data equally well in a
$\chi^2$ sense.  The resulting best-fit models are overplotted on the
data in Figure~\ref{f:bestfit} and described in Table~\ref{t:fitpar}.

The complexity of the dynamical modeling renders direct error
propagation not straightforward in our analysis.  We estimate the
confidence limits using 500 Monte-Carlo realizations of the original
data set.  We analyze each realization using both $\chi^2$ minimization
techniques and find that they yield similar estimates of the
uncertainties.

\begin{table}
  \centering
    \caption{Best-fit model parameters}
    \begin{tabular}{@{}lcc@{}}
    \hline
    Parameter & NGC~2280 & NGC~3810 \\
    \hline
    $V_{c,0}$ (km s$^{-1}$)        & $92 \pm 21$     & $142 \pm 15$    \\
    $\alpha$                     & $0.26 \pm 0.07$ & $0.11 \pm 0.04$ \\
    $\sigma_{R,0}$ (km s$^{-1}$)   & $126 \pm 29$    & $183 \pm 24$    \\
    $\sigma_{z,0}$ (km s$^{-1}$)   & $31 \pm 24 $    & $54 \pm 22 $    \\
    $h_{\rm kin}$ (arcsec)         & $54 \pm 48 $    & $22 \pm 3 $     \\
    $\sigma_z/\sigma_R$          & $0.25 \pm 0.20$  & $0.29 \pm 0.12$ \\
    \hline
    \end{tabular}
    \label{t:fitpar}
\end{table}

\section{Results}
\label{sec:results}

\subsection{NGC~2280}
\label{sec:n2280}

The Scd galaxy NGC~2280 is the latest Hubble type in our full sample of
galaxies with measured velocity ellipsoid axis ratios.  Kinematic data
at radii less than 10 arcsec are dominated by the bulge and bar
components in this system and are therefore excluded from the analysis.
The remaining dataset is limited in its extent but sufficient to derive
a best-fit model, albeit with large uncertainties.  The kinematic scale
length parameter, in particular, has a large uncertainty.  However, this
parameter drops out of the primary quantity of interest, the ratio of
the vertical to radial velocity dispersions.  The best-fit model
velocity ellipsoid axis ratio is $\sigma_z / \sigma_R = 0.25 \pm 0.20$.

\subsection{NGC~3810}

The photometric data for NGC~3810 (see \S\ref{s:phot}) indicates that
the disk light should dominate beyond $R>5$ arcsec in the Sc galaxy
NGC~3810.  However, at these radii the measured velocity dispersions are
not consistent with a stellar disk but instead hint at the presence of a
(hidden) stellar bar.  This is particularly apparent along the major
axis at negative radii where the velocity dispersion profile is flat out
to 8 arcsec.  To avoid the contaminating effect of a potential stellar
bar on velocity ellipsoid modelling we conservatively exclude all
kinematic data at radii less than 10 arcsec.  While this limits the
dataset, its final size and quality are similar to that of the NGC~2280
dataset.  The best-fit model velocity ellipsoid axis ratio is $\sigma_z
/ \sigma_R = 0.29 \pm 0.12$.

\subsection{The Velocity Ellipsoid in Spiral Galaxies}
\label{sec:spirals}

\begin{figure}
   \centerline{\psfig{figure=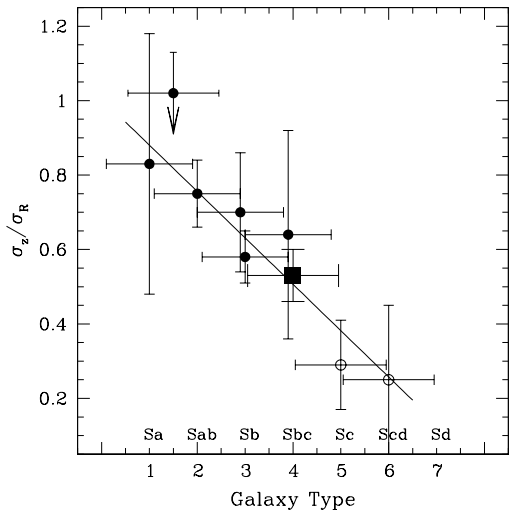,width=0.4\textwidth}}
   \caption{Velocity ellipsoid ratio $\sigma_z/\sigma_R$ as a function
     of galactic type (Hubble stage T).  The solid points show the
     results that we obtained previously \citep{Shapiro2003}. The
     results that we derive in this paper for two late type spirals are
     shown as the open circles.  Horizontal errors represent the
     uncertainty inherent in galaxy classification \citep{Naim1995}.
     The filled square is the value in the Solar Neighbourhood derived
     from Hipparcos data \citep{Dehnen1998}.
   \label{f:fig4}}
\end{figure}

Combining these measurements of \szsr\ for late-type galaxies with our
results for earlier-type spirals (\citealt{Gerssen1997, Gerssen2000,
  Shapiro2003}), we probe the shape of the velocity ellipsoid over the
full extent of the Hubble sequence of disk galaxies.  The results are
shown in Figure~\ref{f:fig4} and reveal a clear trend in \szsr\ with
morphological type.  The combined data are best fit by a line with slope
$-0.12 \pm 0.05$, with a $0.1$\% probability of no correlation.  The
addition of our new results in late-type galaxies confirm and strengthen
the trend identified at much lower significance by \citet{Gerssen2000}
and \citet{Shapiro2003}.

We include in the data and linear regression in Figure~\ref{f:fig4} the
Hipparcos value for the velocity ellipsoid in the solar neighbourhood
($0.53 \pm 0.07$, \citealt{Dehnen1998}), which is consistent with
results based on the RAVE data ($0.55$, \citealt{Casetti2011}, but see
also \citealt{Veltz2008}).  This measurement is obtained at the solar
radius (at $\sim 2-3$ scale lengths in the Milky Way) and is therefore
comparable to our measurements in external galaxies, radially averaged
in each galaxy over a typical range of $\sim 1-3$ disk scale lengths.

Using a different technique, \citet{vdk1999} did not observe a clear
trend in velocity ellipsoid ratio in galaxies of Hubble type Sb to Sd.
However, their results are statistical (the rms of determinations for
multiple edge-on galaxies of the same type) and based on photometry
only.  An advance of this technique combines photometry and spectra of
edge-on galaxies to measure the quantity $\sqrt{M/L}
(\sigma_R/\sigma_z)^{-1}$; assumptions of universal mass-to-light (M/L)
ratios and/or Toomre Q parameters in late-type spirals can then be used
to estimate the velocity ellipsoid ratio \citep{Kregel2005}.  For their
sample of 15 edge-on galaxies of Hubble types Sb to Scd, these authors
obtained a range of estimates of \szsr\ consistent with our direct
measurements and following a marginal trend with Hubble type.

A promising step forward is the use of integral field units (IFUs) to
observe velocity dispersions in disk galaxies.  This is observationally
more efficient than obtaining long-slit spectra, one at a time, along
two (or more) position angles.  Additionally, since IFUs uniformly
sample velocity dispersions along both azimuth and radius across the
disk, the assumption of the epicycle approximation, employed here for
long-slit data, can be relaxed.  \citet{Noordermeer2008} use the PPAK
IFU to explore velocity dispersions in disk galaxies and to constrain
the shape of the velocity ellipsoid in one system, NGC~2985.  Their
measurement of \szsr\ $\approx 0.7$ is consistent with our previous
result for this system ($0.75\pm0.09$; \citealt{Gerssen2000}) and
confirms that the epicycle theory we assume is indeed applicable in this
galaxy.

In a series of conference proceedings \citet{Westfall2008, Westfall2010}
present their ongoing work with the SparsePAK and PPAK IFUs to constrain
the velocity ellipsoid shape and its radial variation in a sample of
spiral galaxies.  These authors find a strong dependence on their
modelled velocity ellipsoid shape with their measurement techniques and
assumptions; they present and describe an analysis comparable to ours
for a single galaxy, NGC~3982 (Sb), for which they find
\szsr\ $=0.31-0.73$ over a radial range of $1-2$ photometric
scalelengths, broadly consistent with our results for galaxies of
similar Hubble type.  These authors also find some evidence for
variation in the velocity ellipsoid ratio with radius, indicating the
potential of IFU data in future studies of the velocity ellipsoid in
external galaxies.

\subsection{The Velocity Ellipsoid across the Hubble Sequence}
\label{sec:ellipticals}

The SAURON team has used the IFU of the same name to study the
kinematics of elliptical and lenticular galaxies and have shown that
many early-type galaxies, the so-called ``fast rotators," are
bulge-dominated galaxies that nevertheless contain a significant disk
component (\citealt{sauronVI, sauronXII}).  It is therefore interesting
to investigate how the disks of these early-type galaxies are related to
those in spiral galaxies.

\begin{figure}
    \centerline{\psfig{figure=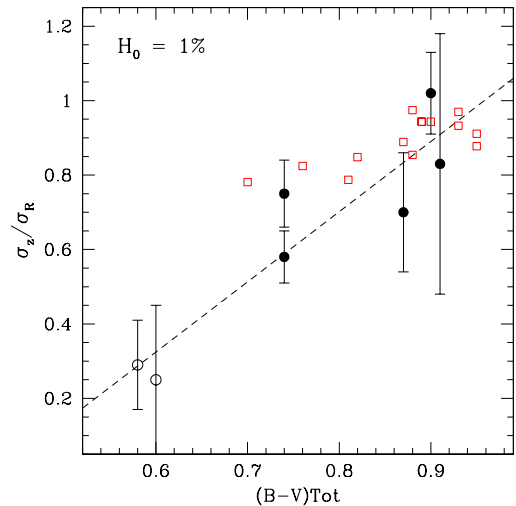,width=0.4\textwidth}}
   \caption{Velocity ellipsoid ratio $\sigma_z/\sigma_R$ as a function
     of inclination- and extinction-corrected galaxy colour (available
     in HyperLeda for 7 of our 8 spirals and the fast-rotating E/S0s).
     The black points indicate our spiral galaxy data, and the red
     squares are the data of \citet{sauronX} for fast-rotating E and S0
     galaxies. The linear fit (dashed line) is to the spiral galaxies
     only. The probability of no correlation, null hypothesis H$_0$, is
     one percent.
   \label{f:fig5}}
\end{figure}

\citet{sauronX} have used axisymmetric Schwarzschild dynamical models to
extract the three-dimensional orbital structure of a subsample of the
SAURON galaxies and measure the shape of their velocity ellipsoids.
These anisotropy measurements are luminosity-weighted, giving more
weight to the high-density equatorial plane, and volume-averaged, giving
more weight to larger radii; as a result, the global anisotropies (see
table 2 of \citealt{sauronX}) are dominated by the disks and are
therefore comparable to our measurements in later-type galaxies
(Cappellari et al. private communication).  Since Hubble T-type is less
meaningful in early-type galaxies than in spirals, we cannot add these
points directly to Figure~\ref{f:fig4}. Instead we use galaxy $B-V$
colour as a proxy for morphological type and plot this quantity against
\szsr\ in Figure~\ref{f:fig5} for both the SAURON sample and our sample
of spiral galaxies.  Colours are taken from
HyperLeda\footnote{http://leda.univ-lyon1.fr/} and are listed for our
sample in Table~\ref{t:samplepar}.

The combined data span the Hubble sequence from E to Scd and show the
strong correlation between velocity ellipsoid ratio and galaxy colour,
as expected from Figure~\ref{f:fig4}, given the known relationship
between galaxy colour and Hubble type.  In Figure~\ref{f:fig5}, we find
a continuous trend of increasing anisotropy in bluer galaxies.
Moreover, we find that the anisotropies of the E/S0 galaxies overlap
with those of the earliest-type spirals in our sample.  However, in the
bluest range ($B-V \approx 0.7-0.8$) of the overlapping data, there is
some evidence that elliptical and lenticular fast rotators are
preferentially more isotropic than their spiral cousins.

\section{Discussion}
\label{sec:discussion}

\subsection{Dependence of the Velocity Ellipsoid Shape on Galaxy Properties}
\label{sec:interpretation}

Figures~\ref{f:fig4} and \ref{f:fig5} point to a deep connection between
galaxy-wide properties and axial ratio of the stellar velocity ellipsoid
in the galaxy disk.  In this and the following sections, we examine the
source of these relationships.

\begin{figure*}
    \centerline{\psfig{figure=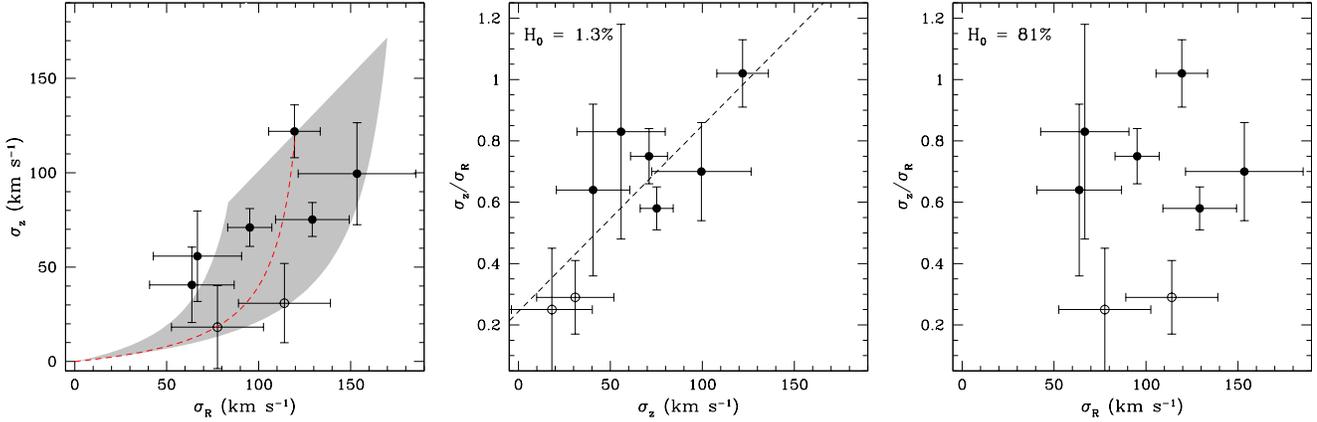,width=\textwidth}}
   \caption{Correlations between the velocity ellipsoid magnitudes and
     shapes for all eight galaxies in our sample.  Here and in
     the following figures, $H_0$ denotes the probability of the null
     hypothesis, i.e. that there is no correlation.  {\bf Left panel:}
     The vertical ellipsoid magnitudes as a function of the radial
     component magnitude.  The dashed line is the projection of the
     best-fit correlation in the middle panel, see text.  The shaded
     area is bounded by the one-sigma uncertainty on the projected
     correlation and by the line \szsr = 1.  {\bf Middle panel:} The
     shape of the velocity ellipsoid for each galaxy in our sample
     plotted as a function of the vertical dispersion magnitude.  The
     results are tightly correlated, dashed line. {\bf Right panel:} The
     velocity ellipsoid shape as a function of the radial dispersion
     magnitude.  There appears to be no relation.
   \label{f:fig6}}
\end{figure*}

The tighter of the two correlations is that between galaxy morphology,
parameterized through Hubble T-type, and the shape of the velocity
ellipsoid (Figure~\ref{f:fig4}).  However, as a semi-quantitative
descriptor of galaxies, T-type is a somewhat subjective property based
on several qualitative criteria that are largely but not perfectly
correlated.  In the de Vaucouleurs RC3 classification used here, these
criteria are the bulge-to-disk ratio and the tightness of the spiral
structure.
Moreover, a number of additional galaxy properties not directly involved
in the determination of T-type also correlate with morphology, such that
earlier-type spirals tend to have higher masses, higher stellar mass
surface densities, redder colours, and lower gas fractions than their
late-type counterparts, albeit with large scatter (see
e.g. \citealt{Roberts1994}).

It is therefore not straightforward to determine which of the
fundamental galaxy properties that correlate with T-type drive the
relation in Figures~\ref{f:fig4} and \ref{f:fig5}.  Broadly, these
properties fall into two general categories: (1) structural properties
related to the large-scale evolutionary history of a galaxy such as
galaxy mass and relative prominence of the bulge, and (2) properties
related to past, current, and future star formation processes such as
galaxy colour, tightness of the spiral structure, and gas fraction.  The
interesting result from Figure~\ref{f:fig4} is that one or both of these
aspects of a galaxy's evolution is intimately related to the heating of
its disk.

Moreover, Figure~\ref{f:fig5} reveals that these same processes are also
at work in fast-rotating early-type galaxies.  The more isotropic shape
of the velocity ellipsoid in early-types, relative to late-types with
similar colours, may indicate a difference in disk heating mechanisms at
work in early- and late-type galaxies.  One possibility is a difference
in strength of a particular mechanism; another is that an additional
heating agent may need to be invoked.

\begin{table*}
  \begin{center}
    \caption{Properties of disk galaxies with velocity ellipsoid shape
      and magnitude measurements.}
    \begin{tabular}{@{}lcccccc@{}}
    \hline
    Galaxy & Hubble Type & $D$ (Mpc) & $\sigma_z/\sigma_R$ & (B-V)Tot & $\Sigma_{\rm{H_2}}$ ($M_\odot$/pc$^{-2}$) & Arm Class \\
    \hline
    NGC 2460  &  Sa      & 20 &    0.83 $\pm$ 0.35 &  0.91  & ...         & 12 \\
    NGC 2775  &  Sa/Sab  & 19 & $<$1.02 $\pm$ 0.11 &  0.90  & 33 $\pm$ 6  & 3  \\
    NGC 2985  &  Sab     & 18 &    0.75 $\pm$ 0.09 &  0.74  & 21 $\pm$ 7  & 3  \\
    NGC 488   &  Sb      & 30 &    0.70 $\pm$ 0.16 &  0.87  & 25 $\pm$ 6  & 3  \\
    NGC 1068  &  Sb      & 16 &    0.58 $\pm$ 0.07 &  0.74  & 41 $\pm$ 14 & 3  \\
    NGC 4030  &  Sbc     & 21 &    0.64 $\pm$ 0.28 &  ...   & 16 $\pm$ 3  & 9  \\
    NGC 3810  &  Sc      & 14 &    0.29 $\pm$ 0.12 &  0.58  & 28 $\pm$ 6  & 2  \\
    NGC 2280  &  Scd     & 26 &    0.25 $\pm$ 0.20 &  0.60  & ...         & 9  \\
    \hline
    \end{tabular}
    \label{t:samplepar}
    \end{center}
\vspace{-0.25cm}
\begin{flushleft}
1) Hubble classifications are from NED. \\
2) Distances are derived from the redshift values listed in NED and assuming a
Hubble constant of $73$ km/s/Mpc. \\
3) Colours are from HyperLeda. \\
4) Arm Class values are from \citet{Elmegreen1987}.
\end{flushleft}

\end{table*}

\subsection{Disentangling Disk Heating Agents}
\label{s:agents}

Of the numerous theoretically studied disk heating agents (see
\S~\ref{sec:intro}), each is expected to behave differently in the
vertical and radial directions as a function of galaxy morphology.  To
constrain the nature of these agents in our spiral galaxies, we
therefore examine the individual axes of the velocity ellipsoid in our
galaxy sample.  (The evolution of the velocity ellipsoid in the SAURON
sample of elliptical and lenticular galaxies is an equally interesting
problem beyond the scope of this paper, and we do not discuss it
further.)

Using the exponential parametrization of \sz\ and \sr\ with
radius (\S~\ref{Modeling}), we evaluate the magnitude of the
dispersion at half the measured kinematic scale length, $h_{kin}/2$.
This radius falls within the range constrained by the observations for
all but one of our eight galaxies.  The local isothermal approximation
of stellar disks predicts that $h_{kin}/2 \approx h_{phot}$
(e.g. \citealp{vdk1986, Martinsson2011}).  Our measurements of $h_{kin}$ are broadly
consistent with this prediction, albeit with large scatter and
noticeable outliers.

The relationship of the magnitudes of the dispersions at $h_{kin}/2$ to each
other and to the shape of the velocity ellipsoid is shown in
Figure~\ref{f:fig6}.  In this figure, the tightest trend, with a 1\%
probability of no correlation, is between the magnitude of the vertical
velocity dispersions \sz\ and the shape of the velocity ellipsoid \szsr;
in stark contrast, there is no correlation between \szsr\ and the
magnitude of the radial velocity dispersions \sr.
Although there is evidence for a relationship between \sz\ and \sr, a
constant ratio (i.e. linear fit) is inconsistent with our direct
measurements of variations in \szsr\ among galaxies and cannot be the
best description of the data.  In the left panel of Figure~\ref{f:fig6},
we therefore plot the prediction for the relationship between \sz\ and
\sr\ based on the observed linear correlation between \sz\ and \szsr\ in the
middle panel.  The projection of the \sz-\szsr\ correlation into the (\sz,\sr)
parameter space is shown as the dashed line.  The uncertainty in this
projection is shown by the shaded area, bounded by the
$1\sigma$ uncertainties on the linear fit and by the line where
\szsr$=1$.  Within the scatter, the predicted relationship between
\sz\ and \sr\ matches the data.

These results are a powerful new tool in probing the nature of disk
heating agents at work across the Hubble sequence, and they demonstrate
the feasibility of observationally disentangling disk heating agents in
spiral galaxies.  In particular, Figure~\ref{f:fig6} places two
important constraints on disk heating mechanisms:

(1) There is a disk heating agent that is responsible for increasing
both vertical and radial dispersions, which we call the
``three-dimensional agent."  The trend of \sz\ with \sr\ demonstrates
that, in external galaxies, both dispersion components grow
simultaneously although not at the same rate, as is also observed in the
solar neighbourhood (e.g. \citealt{Holmberg2007}).  The tight correlation
between \sz\ and \szsr\ additionally requires that the three-dimensional
agent responsible for the magnitude of the vertical dispersions also
determine the shape of the velocity ellipsoid, such that larger vertical
dispersions are coincident with more isotropic heating and smaller
vertical dispersions are coincident with more radially anisotropic
heating.  The correlations in Figures~\ref{f:fig4} and \ref{f:fig6}
therefore imply that the strength of the three-dimensional
agent in a galaxy should correlate strongly with \sz, \szsr, and the
galaxy properties associated with T-type and correlate weakly with \sr.

(2) It is likely, given the observed scatter in the \sz-\sr\ relation, that
there is also a second agent involved in radial heating, which we call
the ``radial agent."  The lack of correlation between \sr\ and
\szsr\ suggests that this heating mechanism has no effect on the vertical
dispersions (and therefore on the shape of the velocity ellipsoid) and
operates only in the plane of the galaxy.  Additionally, the lack of
correlation between \sr\ and \szsr\ implies that the presence or
strength of the radial agent in a galaxy should correlate only with \sr,
and not with \sz, \szsr, or the galaxy properties associated with
T-type.

Two of the most commonly invoked disk heating agents are GMCs and
transient spiral structure (\S~\ref{sec:intro}), with the former
expected to operate in three dimensions and the latter only in the plane
of the disk.  In Sections~\ref{sec:gmcs} and \ref{sec:armclass}, we
investigate whether these heating mechanisms are consistent with our
observational constraints on the three-dimensional agent and radial
agent, respectively.

\subsection{GMCs as a Heating Agent}
\label{sec:gmcs}

We estimate the molecular gas surface mass densities for six of the
eight galaxies in our sample for which the CO models of
\citet{Young1995} are available.  We compute $\Sigma_{H_2}$ in each
system as the average surface density over the same range in radii used
in our velocity ellipsoid measurements.  The results, with rather large
error bars, are tabulated in Table~\ref{t:samplepar}.

\begin{figure*}
    \centerline{\psfig{figure=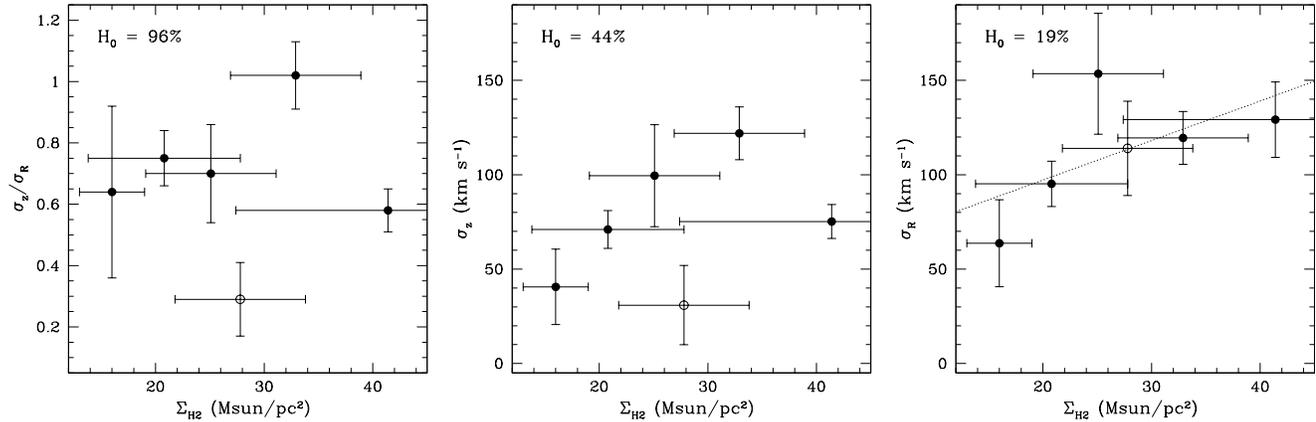,width=\textwidth}}
   \caption{The velocity ellipsoid shape and magnitudes as a function of
     the H$_2$ gas surface density. Gas densities are estimated from the
     CO measurements of \citet{Young1995} and are computed as average
     values over the radii used in our kinematic analysis.  {\bf Left:}
     The velocity ellipsoid shapes are not correlated with the molecular
     gas surface density.  {\bf Middle:} There is a hint that the
     vertical component of the velocity dispersion is correlated with
     $\Sigma_{H_2}$, but the scatter is too large to state this
     conclusively. {\bf Right:} The radial component increases with
     molecular gas density.
   \label{f:fig7}}
\end{figure*}

In Figure~\ref{f:fig7}, the velocity ellipsoid shape and magnitude are
shown as a function of $\Sigma_{H_2}$.  If molecular gas is the dominant
``three-dimensional agent'', we expect the gas surface density to
correlate with \sz, \sr (weakly), \szsr, and T-type.  The radial
dispersion shown in the right panel are indeed consistent with the
expected increase in dispersions with gas surface density; however, a
trend is not evident in the vertical dispersions (middle panel).
The probability that neither radial nor vertical dispersions are
correlated with surface density is 19\% and 44\%, respectively.
Moreover, $\Sigma_{H_2}$ is also not correlated with the velocity
ellipsoid shape \szsr\ and is instead consistent with scattering around
a constant value. 
\citealt{Roberts1994} show that the surface density of molecular gas
increases with Hubble T-type, although this trend is not clearly
reflected in our much smaller sample (not shown).  Together, these
relationships point to GMCs as only marginally consistent with being the
three-dimensional agent.

Theoretically, three-dimensional heating via GMCs has been shown to
drive the velocity ellipsoid axis ratio from an initially isotropic
distribution to a small equilibrium value (e.g. \citealt{Jenkins1990}).
Interestingly, \citet{sellwood2008} argued that this mechanism
alone can explain the observed disk heating in the Milky Way, with the
degree of anisotropy in the distribution of the scattering events
determining the final equilibrium ratio \szsr.  However, our observed
variation in \szsr\ with T-type in external galaxies is larger than can
be comfortably explained with GMCs as the only heating mechanism.

The identification of GMCs as the three-dimensional heating agent is not
straightforward.  GMCs are consistent with the behavior required by the
data but leave room for alternative explanations.  \citet{Jenkins1992}
likewise notes that the observations of constant disk scale height with
radius in external galaxies is difficult to reconcile with the observed
exponential distribution of molecular clouds if GMCs are important disk
heating agents.  Other potential three-dimensional agents include
transient asymmetric spiral structure, dark halo objects (black holes
and dark matter substructure), globular clusters, and accreting
satellite galaxies
\citep{Saha2010,Lacey1985,Hanninen2002,vandePutte2009,Velazquez1999}.
Of these, numerical simulations have demonstrated that dark halo objects
and globular clusters are unlikely to exist in sufficient numbers to
produce the observed heating
\citep{Lacey1985,Hanninen2002,vandePutte2009, Benson2004}, leaving
spiral structure and accreting satellites as viable mechanisms.  It is
not obvious how the accretion history of a galaxy might be quantified
and subsequently compared to the observed shape of the velocity
ellipsoid, so observational tests of the viability of this alternative
heating mechanism are not straightforward.  In contrast, spiral
structure is a direct observable, and we examine the effects of this
potential heating agent in the following section.

\begin{figure*}
    \centerline{\psfig{figure=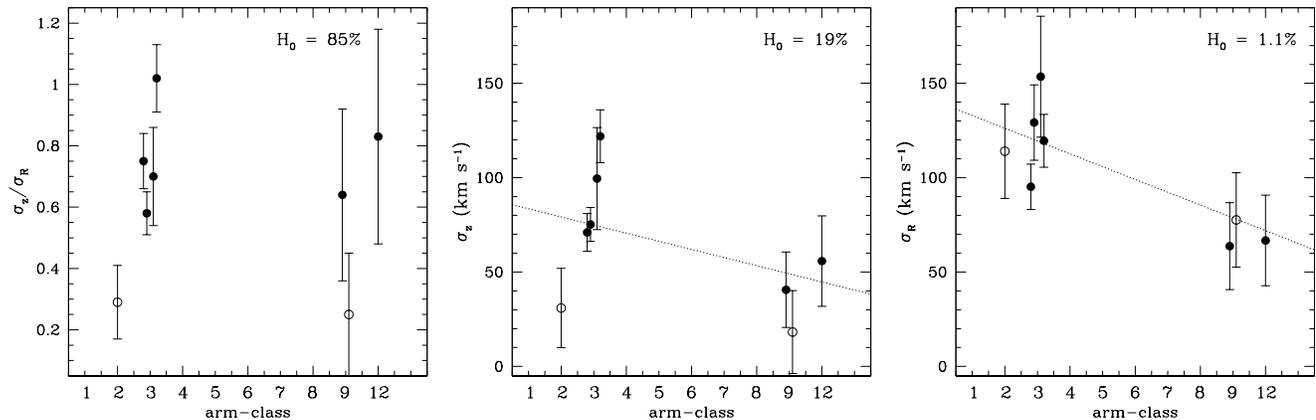,width=\textwidth}}
   \caption{The velocity ellipsoid shape and magnitudes as a function of
     arm class, as defined in \citet{Elmegreen1987} to quantify the
     orderliness of spiral structure from flocculent (class 1) to
     grand-design (class 12).  Note that there is no arm-class 10 and 11
     (cf., Figure~1 in \citealt{Elmegreen1987}).
     {\bf Left:} The velocity ellipsoid shape is not correlated with arm
     class.  {\bf Middle:} The vertical magnitude of the ellipsoid
     decreases with arm class.  {\bf Right:} There is a clear trend
     between arm class and radial component, as expected from the Toomre
     $Q$ criterion, see text. Note that for plotting purposes we have
     added small offsets to galaxies with the same arm class.
   \label{f:fig8}}
\end{figure*}

\subsection{Spiral Structure as a Heating Agent}
\label{sec:armclass}

To investigate the role of spiral structure on disk heating we use the
arm class parameter, developed by \citet{Elmegreen1987}, to quantify the
strength and symmetry of the spiral structure in galaxies.  They define
this parameter to describe the orderliness of spiral structure from
flocculent (class 1) to grand-design (class 12). The arm class
parameters for our sample are listed in Table~\ref{t:samplepar}.

We compare the velocity ellipsoid shape and magnitudes in our sample
galaxies to their arm classes in Figure~\ref{f:fig8} and find a strong
correlation between \sr\ and arm class.  This can be interpreted as
spiral structure preferentially forming in Toomre unstable disks (Toomre
$Q$ stability criterion; \citealt{Toomre1964}).  The Toomre $Q$ is
proportional to \sr, such that larger velocity dispersions create more
stable systems, which are then less responsive to the formation of
(transient) spiral structure \citep[e.g.][]{Binney2012}.  In our sample,
we indeed find the most clear spiral structure (class 12) in those
galaxies with the lowest \sr.

In Figure~\ref{f:fig8}, there is a marginal relationship between
\sz\ and spiral structure at much lower significance than that with \sr,
and no trend is evident between \szsr\ and arm class.  This latter
result is consistent with the absence of a trend between arm class and
T-type found by \citet{Elmegreen1987} and confirmed in our much smaller
sample (not shown).  Spiral structure thus correlates primarily with
\sr, weakly (if at all) with \sz, and not with \szsr\ nor T-type.  Based
on the requirements of different heating agents described in
Section~\ref{s:agents}, spiral transients are excellent candidates for
the ``radial agent".  

\section{Conclusions}
\label{sec:conclu}

In this paper, we present the kinematics and modelled velocity ellipsoid
shape, i.e. the ratio of the vertical and radial velocity dispersions
\szsr, in two late-type disk galaxies, NGC~2280 (Scd) and NGC~3810 (Sc).
Combining these results with our previous measurements of this ratio in
six early type disks demonstrates that the shape of the velocity
ellipsoid is strongly correlated with Hubble type and with galaxy color.
Moreover, early-type galaxies (E/S0) with significant disk components
appear to follow a related trend.

Examination of the relationship of vertical and radial dispersions
(\sz\ and \sr) with each other and with \szsr\ shows that
\szsr\ correlates strongly with \sz\ but not at all with \sr.  This
suggests that kinematic observations alone can disentangle the effects
of disk heating mechanisms that act in three-dimensions from the effects
of those that act predominantly in the plane of the disk.

Two commonly invoked disk heating mechanisms are stellar scattering off
of giant molecular clouds (as the three-dimensional agent) and
perturbations from transient spiral structure (as the radial agent).  To
test the applicability of these theories to our data across the full
range of morphological types, we probe the relationship of molecular gas
surface mass density and of spiral arm structure to our observed
dispersions (\sz\ and \sr) and their ratio (\szsr).  The data show that
spiral structure varies between galaxies exactly as expected of the
radial heating agent, and therefore it seems likely that spiral
structure is responsible for (morphologically-independent) heating in
the plane of galactic disks.  The data are tentatively consistent with
GMCs being the three-dimensional agent, although the gap between the
expected and observed variation between galaxies for the
three-dimensional agent leaves room for alternative interpretations.

These results suggest exciting possibilities for future work in directly
probing disk heating agents in external galaxies.  Open questions
include the nature of the three-dimensional agent and its relationship
to the molecular content of galaxies, the possible relationship of
stellar populations to disk heating mechanisms, and the relationship of
disk heating mechanisms in fast-rotating early-type galaxies to those of
spirals.

\section*{Acknowledgments}

We thank the staff of the La Silla Observatory, in particular Emanuela
Pompei, for their outstanding support during our observing runs.  This
paper also benefited significantly from discussions with Anne-Marie
Weijmans, Cecilia Scannapieco, Ivan Minchev and Jes\'us Falc\'on
Barroso.  We also thank the referee for helpful comments.  KSG wishes to
acknowledge the Sigma Xi Grant in Aid of Research for travel support
that enabled the observations presented here.  This research has made
use of the Hyperleda database (http://leda.univ-lyon1.fr) and of the
NASA/IPAC Extragalactic Database, which is operated by the Jet
Propulsion Laboratory, California Institute of Technology, under
contract with the National Aeronautics and Space Administration.

\bibliography{biblio}
%\bsp

\label{lastpage}

\end{document}